\documentclass[pre,twocolumn,showpacs,superscriptaddress]{revtex4-1}  
\usepackage{natbib}
\usepackage{graphicx}
\usepackage{subfigure}
\usepackage{epsfig} 
\usepackage{amsmath}

\usepackage{color}
\newcommand{\red}[1]{\textcolor{black}{#1}}

\begin{document}

\title{Position-dependent radiative transfer as a tool for studying Anderson localization:\\
Delay time, time-reversal and coherent backscattering}

\author{B.A. van Tiggelen}
\affiliation{Universit\'{e} Grenoble Alpes, LPMMC, F-38000 Grenoble, France}
\affiliation{CNRS, LPMMC, F-38000 Grenoble, France}
\author{S.E. Skipetrov}
\affiliation{Universit\'{e} Grenoble Alpes, LPMMC, F-38000 Grenoble, France}
\affiliation{CNRS, LPMMC, F-38000 Grenoble, France}
\author{J.H. Page}
\affiliation{Department of Physics and Astronomy, University of Manitoba, Winnipeg, Manitoba R3T 2N2, Canada}

\date{\today}

\begin{abstract}
Previous work has established that the localized regime of wave transport in open media is characterized by a position-dependent diffusion coefficient. In this work we study how the concept of position-dependent diffusion affects the delay time, the transverse confinement, the coherent backscattering, and the time reversal of waves. Definitions of energy transport velocity of localized waves are proposed. We start with a phenomenological model of radiative transfer and then present a novel perturbational approach based on the self-consistent theory of localization. The latter allows us to obtain results relevant for realistic experiments in disordered quasi-1D wave guides and 3D slabs.
\end{abstract}


\maketitle

\section{Introduction}

Among many other features, Anderson localization of waves is characterized by a halt of diffuse transport \cite{PT}. Since the formulation of the scaling theory of localization we understand that diffusion cannot entirely vanish in open media due to leakage of waves across the sample boundaries \cite{g4}. The result is a suppressed but scale dependent conductance, depending in a universal way on the size and dimensionality of the random medium. Later work established that many features of scaling theory can be understood from the self-consistent theory of localization \cite{sc}, which adopts the constructive interference of time-reversed waves as the sole mechanism of the suppression of diffusion. More recently, this theory was extended to predict a position-dependent diffusion constant \cite{prl2000,dz,dz2}. Microscopic derivations were provided using diagrammatic \cite{dz3} and super-symmetric \cite{dz4} approaches. Diffusion is suppressed deep inside the sample, yet hardly near the boundaries. This result is physically plausible, and consistent with scaling theory for macroscopic transport quantities such as, e.g., the conductance. It was tested against numerical simulations \cite{payne2010} and observed in an experiment \cite{yamilov2014}.

If stationary transport is described by a spatially varying local diffusion constant, what does this imply for the dynamics, and in particular for the energy transport velocity $v_E$? In a weakly disordered three-dimensional (3D) medium the wave transport is diffusive with the diffusion constant given by $D = v_E \ell/3$ \cite{vE}. It was shown that $v_E$ is intrinsically a dynamic property, whereas the transport mean free path $\ell$ emerges by itself in stationary (DC) measurements, like in diffuse transmission through a slab of thickness $L$, $T \sim \ell/L$. This is important because the transport velocity can be very small due, for example, to strongly resonant scattering, and can thus lead to ``small'' diffusion constants. It can be then difficult to distinguish between situations in which $D$ is small due to Anderson localization effects leading to small $\ell$ or due to small $v_E$. Up to now, transport theories of localization essentially concentrated on $\ell$ and not on $v_E$. So far we know that $v_E$ depends on many sample properties such as, e.g., the scattering crosssection of scatterers and their number density, but not on sample size or boundary conditions. Is $v_E$ well defined in the localized regime? What does the position-dependent $\ell$ imply for the transverse spreading of a wave packet in experiments similar to those of Ref.\ \cite{john}?  Similar questions arise for the Wigner delay time in reflection and transmission, and for coherent backscattering (CBS). In this work, we first develop general arguments for the delay time in a medium with position-dependent $\ell$, valid under very broad conditions, and then present a perturbational approach to Anderson localization in the framework of the self-consistent theory of localization. The latter is applied to study wave dynamics, the transverse spreading of a wave packet, CBS, and time-reversal of localized waves.

\section{Friedel identity in radiative transfer}
\label{friedelsec}

Depth-dependent extinction is very common in radiative transfer. Here we  \emph{conjecture} that the phenomenological equation of radiative transfer (EQRT) applies in the localized regime though with a depth-dependent scattering mean free path $\ell(z)$. This is clearly an oversimplified picture. In particular it disregards off-shell scattering ($\omega \neq kc$)  that becomes significant when the scattering mean free path becomes small. Nevertheless, it is consistent with the macroscopic picture of depth-dependent diffusion.
In this section we make the connection between phase delay time, a highly mesoscopic wave property, and the EQRT, apparently a classical equation where all wave properties seem to have disappeared. The essential elements in this approach are the transport velocity and the link between stored energy and delay time established in condensed matter physics. We will reproduce some exact results from standard radiative transfer theory, such as the relation between incoming flux, source and energy density away from the boundaries, that will be necessary to elucidate the exact role of the incident flux for the delay time, and its scaling with the sample size.

We consider a slab of disordered medium confined between the planes $z=0$ and $z=L$ and made of isotropic, conservative scatterers. The incoming specific intensity  is $I(0,  0< \mu <1,s) $ on the left and $I(L,  -1 < \mu <0,s) $ on the right. The EQRT can be written as \red{\cite{chandra1950,ishimaru1978,akker2007}}
\begin{eqnarray}\label{EQRT}
   \frac{s}{v_E} I(z,\mu,s) &+& \mu \partial_z I(z,\mu,s) + \frac{1 }{\ell(z)}I(z,\mu,s) \nonumber \\
   &= & \frac{1}{\ell(z)} \frac{1}{2}\int_{-1}^1 d\mu' I(z,\mu',s),
\end{eqnarray}
where $s$ is the Laplace conjugate of time and $\mu = \cos \theta$. All observables in this paper are ensemble-averaged and if no confusion exists, no explicit reference to this will be given. We have assumed the existence of an  energy velocity $v_E$ independent of the direction of scattering $\mu$ and position (or depth)  $z$. We can introduce the optical depth $d\tau = dz/\ell(z)$ and write
\begin{eqnarray}\label{EQRT2}
   \frac{s}{v_E} \ell(\tau) I(\tau,\mu,s) &+& \mu \partial_\tau I(\tau,\mu,s) + I(\tau,\mu,s) \nonumber \\
   &= & \frac{1}{2}v_E w(\tau,s)
\end{eqnarray}
with the  energy density $w(\tau,s) = v_E^{-1}\int_{-1}^1 d\mu  I(\tau,\mu,s)$ (and equal to $2J/v_E$ in terms of the source function $J$ featuring in radiative transfer).

Let us first obtain a useful result relevant to the study of the delay time. Upon integrating Eq.\ (\ref{EQRT2}) over the depth $z$ and over all angles we see that
\begin{eqnarray}\label{store1}
    sW(s) + \left[ F^+(L,s) \right. &+& \left. F^-(0,s) \right] \nonumber \\
    &=& \left[F^-(L,s) + F^+(0,s)\right].
\end{eqnarray}
The term in square brackets on the left-hand side contains transmitted and reflected fluxes $F^+(L,s)$ and $F^-(0,s)$, respectively, the right-hand side is the incident flux on both sides of the slab. The total energy is $W = S \int_0^B d\tau \ell(\tau) w(\tau) = S\int_0^L dz w(z)$, with the total optical thickness $B=\tau(L)$ and slab surface $S$. To discuss the delay time, we assume the incident flux independent of $s$ (i.e., perfect delta functions in time): $F^-(L,s) + F^+(0,s) = F_{in}$. The average transmission and reflection coefficients are then $T(s)=F^+(L,s)/F_{in}$ and $R(s)=F^-(0,s)/F_{in}$, respectively. For $s=0$ (time-integrated signal), we infer flux conservation $R(0)+T(0)=1$. Taking the derivative of Eq.\ (\ref{store1}) with respect to $s$ we obtain
\begin{eqnarray}\label{store2}
    \frac{W(s=0)}{F_{in}} &=&  -\left. \frac{dT(s)}{ds} \right|_{s=0} - \left. \frac{dR(s)}{ds} \right|_{s=0} \nonumber \\
    &=& \left\langle T(\omega) \frac{d\phi_T(\omega)}{d\omega} \right\rangle + \left\langle R(\omega) \frac{d\phi_R(\omega)}{d\omega} \right\rangle,\;\;
\end{eqnarray}
where the angular brackets $\langle \cdots \rangle$ indicate that the ensemble averaging is to be carried out for the products $T d\phi_T/d\omega$ and $R d\phi_R/d\omega$.
This equation makes the desired connection between stored energy and total (channel-summed) phase delay time. The second equality follows from the notion that the complex transmission coefficient is $t= \sqrt{T}\exp(i\phi_T)$, with $\phi_T(\omega)$ being the phase shift, and that $\frac{1}{2}\ln T + i\phi_T$ is a function of $\frac12 s + i \omega$, so that the Cauchy-Riemann equations give $d \phi_T(\omega)/d\omega = -d\ln T(s)/ds$ at $s=0$ (and similarly for the reflection coefficient). The relation between stored energy (or charge) and phase delay time is well known in different contexts: as Friedel's identity \cite{mahan} in the context of screening of charge around impurities, or as Jauch formula  relating phase delay time to the local density of states in scattering theory \cite{jauch}; \red{see also Ref.\ \cite{akker12} for a related discussion and a list of relevant references.} In this case the relation between stored energy in radiative transfer and the mesoscopic density of states is controlled by the transport velocity $v_E$, which, for the moment, appears  just phenomenologically in the EQRT. The phase delay time, in turn, can easily be related to the first moment of the scattered intensity, i.e.  $\int_0^\infty dt T(t) t  = \langle T(\omega) d\phi_T(\omega)/d\omega \rangle$, and similarly for reflection, which explains its interpretation as average delay time.

Let us next consider the stationary energy flow putting $s=0$  and agree to have unit incident flux. Upon integrating Eq.\ (\ref{EQRT2}) over angles we obtain the following equation for the energy density:
\begin{eqnarray}\label{WW}
    w(\tau) = S_L (\tau) &+& S_R(B-\tau) \nonumber \\
    &+& \frac{1}{2} \int_0^{\tau} d\tau' E_1(|\tau-\tau'|) w(\tau'),
\end{eqnarray}
This identifies the  source in radiative transfer as  $S_L (\tau) =v_E^{-1}\int_0^1 d \mu I(0,\mu)\exp(-\tau/\mu)$ in terms of the incident radiation, and similarly for $S_R$ on the other side of the slab. Note that
 \begin{equation}
    \int_0^\infty d\tau S_L (\tau)  = \frac{f^+(0)}{v_E},
 \end{equation}
 that is, the integral over optical depth of the energy source is determined  by the incident flux density $f^+= F^+/S$.

A useful identity can be obtained by defining $K (\tau) \equiv \int_{-1}^1 d\mu \mu^2 I(\tau, \mu)$. We easily see that $\partial_{\tau} K = -f$. Since the total flux is conserved, $K(\tau) = -f\tau + \mathrm{const}$. We can check that this imposes the  radiation $I_\infty - \frac{3}{2}f(\tau-\mu)$ far from the boundary, with
 \begin{equation}\label{KF}
    \frac{2}{3}I_\infty = \int_{0}^1 d\mu \mu^2 I(0, \mu) +\int_{0}^1 d\mu \mu^2 R(\mu).
 \end{equation}
This relation connects the incident radiation on the surface, the reflected intensity and the one in the interior of the sample. For a conservative half space, $f=0$, and the radiation pattern reaches the isotropic intensity $I_\infty$. The isotropic incident radiation $I(0,\mu> 0) = 2 f_{in}$ has incident flux density $ f_{in}=  F_{in}/S$, and Eq.~(\ref{KF}) immediately gives us the (expected) result that $I_\infty= 2 F_{in}/S$, and hence the constant energy density $w= 4F_{in}/Sv_E$ away from the boundaries.

Previous work  \cite{blanco} showed  that the delay time is essentially determined by a typical geometric length scale of the medium and largely independent from scattering details of bulk and surface, \emph{provided the incident radiation is isotropic}. Pierrat et al. \cite{pierrat}  confirmed this observation but emphasized the persisting role of transport velocity. Following this previous work, we can write for the total \emph{channel-averaged} delay time $\langle t \rangle = \langle T(t) t \rangle + \langle R(t) t \rangle$ for an optically thick medium of arbitrary geometry with volume $V$ and boundary surface $S$,
 \begin{equation}\label{tnu}
   \langle t \rangle  = \frac{4 \nu}{v_E}  \frac{V}{S},
 \end{equation}

For isotropic radiation incident on a slab we just derived that $w= 4F_{in}/Sv_E$ away from the boundaries.
This result looks ``universal" and likely to be valid in more general geometries. Hence, by neglecting the surface layer, $\langle t \rangle = W/F_{in}= 4V/Sv_E$ which confirms the universal value  $\nu = 1 $ reported in Refs.\ \cite{blanco,pierrat}. We recall that the energy transport velocity so far only appears phenomenologically in the dynamics of the EQRT, without any microscopic interpretation in terms of scattering properties. We can now provide a more microscopic definition.  Equation~(\ref{store2}) states that $\langle t \rangle = W/F_{in}$. Combining this with Eq.~(\ref{tnu}) yields for the energy transport velocity
 \begin{equation}\label{vE}
    v_E = \frac{4f_{in}}{\overline{w}},
 \end{equation}
 where, given isotropic incident radiation,  $\overline{w} = W/V$ is the volume-averaged stored energy density in the medium and $f_{in}$ the incident flux density assumed constant across the boundary surface $S$. Because all quantities on the right-hand side are well defined and measurable, this equation can actually serve as the \emph{definition} of the energy transport velocity, even in the localized regime. This definition is much in the spirit of the ``energy velocity" as defined by Loudon \cite{loudon}. It is also clear that via $\overline{w}$ the energy velocity becomes connected to the density of states in the medium. In the next section, we will find that, even in the localized regime with scale-dependent diffusion, $W$ scales like the volume $V$ so that  $\nu$ is scale-invariant though not always equal to $1$ if the incident radiation is not isotropic.

\section{Examples of delay time calculations}

The most  remarkable aspect of Eq.~(\ref{store2}) is that total delay time---an intrinsic dynamic quantity---is related to the stored energy that can be found from the stationary EQRT (i.e. putting $s=0$). In this section we will calculate total delay under different conditions.  In the next section we will address delay in reflection and transmission separately, and will see that the dynamics then explicitly comes in.

 \subsection{Delay for isotropic incidence}

 For isotropic incident  radiation equal on both sides of the slab  we expect $f=0$ and $K(\tau)$ to be constant throughout the slab. Hence $I$ is constant and isotropic everywhere.  The main mathematical reason for this almost trivial result is that the orientational and depth dependencies of the specific intensity are strongly connected by EQRT. Absence of the first implies absence of the second.

 Isotropic radiation $I$ everywhere leads to  constant energy density  $w =2 I/v_E$. In its turn, $I   = v_E w/2 $ implies a flux $F_{in}^\pm = \frac{1}{4} S v_E w $ incident on both sides at the surface $S$. Since $W=w \times LS = w\times V $, the channel-averaged delay time is
 \begin{equation}\label{iso}
   \langle t \rangle = \langle T(t) t \rangle + \langle R(t) t \rangle  = \frac{2 L }{v_E}= \frac{4 V}{ S_{tot} v_E}.
 \end{equation}
 The result $\langle t \rangle = 2L/v_E$ holds for the slab geometry but it is clear that the argument of constant energy density is valid for any geometry with a total surrounding surface  $S_{tot}$ ($= 2S$ for a slab) and a volume $V$, and even if the waves are localized by disorder. Hence $\nu=1$ in Eq.\ (\ref{tnu}) and we recover the somewhat counterintuitive result by Blanco and Fournier \cite{blanco}.

If the number density $n$ of the scatterers is small enough, we expect ${W} = w_0V +  N W_S$, where $W_S$ is the total energy stored inside each scatterer, given the constant energy density $w_0$ outside. If $v_p$ is the phase velocity in the medium, we can write \cite{vE} $f_{in} = I/2= \frac{1}{4 }w_0 c_0^2 k/\omega = w_0c_0^2/4v_p$ and $\overline{w} = w_0+nW_S $. Thus, $1/v_E = v_p/c_0^2 \times 1 + nW_S/w_0$, which is the microscopic result, and scale independent. If the scatterer density is high enough for the waves to be localized, this result no longer applies. Nevertheless, $W$ still scales with the volume $V$ so that  the definition (\ref{vE})  for  $v_E$ does not reveal any scale dependence. A hand waving argument could have given $W \sim \xi^3$ (with $\xi$ the localization length) rather than $W \sim L^3$. The dependence of energy density on depth $w(z)$  will be discussed in the next section and will help to explain why this argument is wrong.

 \subsection{Delay for normal incidence}

For the incident wave normal to the surface $z=0$ of a thick slab one finds the stationary reflection coefficient $R(\mu) =\frac{1}{2}\sqrt{3} H(1,\mu)$, which carries a unit flux density $f^-(0) = \int_0^1 d \mu \mu R(\mu)$. At the same time, $\int_{0}^1 d\mu \mu^2 R(\mu) =  \tau_0$, with $\tau_0$ the extrapolation length in units of the mean free path \cite{henk}. Equation (\ref{KF}) then tells us that $ I_\infty= \frac{3}{2} (1+\tau_0)$ thus $I(\tau, \mu) =\frac{3}{2} \left(1+\tau_0-f\tau +f\mu\right)$. Hence $w(\tau)=
3\left(1+\tau_0-f \tau \right)/v_E$. If we  neglect the small energy contained in the boundary layers, we can impose $W(B)=0$ so that the total flux density is $f=(1+\tau_0)/B$.  From the Friedel identity, the total, channel-averaged delay time is
\begin{eqnarray}\label{normal}
\langle t \rangle &=& \langle T(t) t \rangle + \langle R(t) t \rangle
\nonumber \\
&=& 3(1+\tau_0)\frac{1}{v_E} \int_0^L dz \left(1 - \frac{\tau}{B} \right).
\end{eqnarray}
 Recall that $dz = d\tau \ell(\tau) $ and that $\ell(\tau) =\ell(B-\tau)$ if we assume that boundary conditions are identical on both sides of the slab. Thus,
 \begin{eqnarray}\label{normal2}
\langle t \rangle  &=& \frac{3}{2}(1+\tau_0)\frac{1}{v_E} \int_0^L dz \left(1 - \frac{\tau}{B} + \frac{\tau}{B} \right) \nonumber \\ &=&
 \frac{3}{2}(1+\tau_0)\frac{L}{v_E}.
 \end{eqnarray}
This result is the exact outcome of radiative transfer theory for normal incidence on a slab without internal reflections ($\tau_0=0.7104\ldots$), for an \emph{arbitrary} (symmetric) profile $\ell(z)$. We find here $\nu =3(1+\tau_0)/4 \approx 1.28$ in Eq.~(\ref{tnu}).

 \subsection{Delay in the diffusion approximation}

 In the diffusion approximation (DA) we replace EQRT (\ref{EQRT}) by the following diffusion equation:
 \begin{eqnarray}
\left[  \frac{s}{v_E} \right. &+& \left. \frac{1}{3}\ell(z, s)q^2 \right.
\nonumber \\
&+& \left. \frac{1}{3}\partial_z\ell(z, s)\partial_z \right] G(z,z',s,q)  =  \delta(z-z').
\label{da}
\end{eqnarray}
Here, $G(z,z',s,q)$ is the Fourier-Laplace transformation of $G(z,z',t,\mathbf{R})$ which stands for the energy density at time $t$, depth $z$ and transverse distance $\mathbf{R}$, given a source at $t'=0$, $z'$, and $\mathbf{R}'=0$. The solution for the energy density given an arbitrary incident radiation $I(0,\mu > 0,s,q)$ (where the $q$-dependence determines the transverse profile of the incident beam) is $w(z,s,q)= \int dz_S G(z,z_S,s,q) S(z_S,s,q) $ where we recall from the previous section that the source is  $S(z,s,q)= (1/v_E) \int_0^1 d \mu I(0, \mu, s, q) \exp(-\tau\mu)$. Upon substituting $d\tau = dz/\ell(z,s)$, the diffusion equation takes the form
\begin{eqnarray}
\left[ \frac{s}{v_E}\ell(\tau,s) \right. &+& \left. \frac{1}{3} \ell(\tau,s)^2 q^2  \right.
\nonumber \\
&-& \left. \frac{1}{3}\partial^2_\tau \right] G(\tau,\tau')=  \delta(\tau- \tau'),
\label{da2}
\end{eqnarray}
where we omitted $s$ and $q$ as explicit arguments of $G$ for brevity.
For the total delay time we just need $s=0$. We shall use simplified  boundary conditions $G = 0 $ at $\tau = B+\tau_0 $ and $ \tau =-\tau_0$ with  $B = \int_0^L dz \, /\ell(z)$ the total optical thickness. This allows us to avoid mixed boundary conditions but generates small errors of order $\tau_0/B$ in the energy density.  For $s=0$ and $q=0$ the eigenfunctions are $ \Phi_n(\tau) = \sqrt{2/B^*} \sin[q_n(\tau+\tau_0)]$ with eigenvalues $q_n = \pi n /B^*$ (we shall write $B^*=B+2\tau_0$ and $\tau^* = \tau+\tau_0)$, and the solution of Eq.\ (\ref{da2}) is
\begin{equation}\label{Gserie}
  G(\tau, \tau') = \frac{2}{B^*} \sum_{n=1}^\infty \frac{\sin(q_n \tau^*)\sin(q_n\tau'^*)}{\frac{1}{3} q^2_n}.
\end{equation}

The stored energy can be calculated as
\begin{eqnarray}
  W(s) &=& \int d^2 \mathbf{r} \int_0^L dz \,  w(z,\mathbf{r},s) \nonumber \\
  &=& (2\pi)^2 \frac{S}{v_E} \int_0^L dz \, G(\tau, \tau_S, s, q=0)
  \nonumber \\
  &\propto& \int_0^L dz \, W(\tau, s)
  \label{ws}
\end{eqnarray}
with $W(\tau, s) \propto v_E^{-1}G(\tau, \tau_S, s, q=0)$.
We shall ignore the front factor that drops out in the delay time.  Upon inserting the eigenfunction expansion (\ref{Gserie}) for $G$ and assuming $\tau_S \ll B$, we obtain
\begin{eqnarray}\label{Gserie2}
W(\tau, s=0) &=& \frac{6B^*}{v_E\pi^2 } \sum_{n=1}^\infty \frac{\sin(q_n\tau_S^*)\sin(q_n\tau^*)}{ n^2}\\
  \, &\approx&  \frac{6\tau_S^*}{v_E\pi  } \sum_{n=1}^\infty \frac{\sin(q_n\tau^*)}{n} \\ &=&
  {\frac{3}{v_E}\tau_S^*}(1-\tau^*/B^* ).
\end{eqnarray}
And, finally,
\begin{eqnarray}
  W &=& W(s=0) = \int_0^L dz \, W(\tau, s=0) \nonumber \\
  &=&  \frac{3}{v_E}(\tau_S+\tau_0)\int_0^L dz [1-\tau(z)^*/B^* ].
  \label{totalw}
\end{eqnarray}
This result resembles closely Eq.\ (\ref{normal}) obtained from the radiative transfer theory. Note that we have assumed nothing yet about $\ell(z)$. The last simplification can be made if we assume that $\ell(z) = \ell(L-z)$ which is true in the self-consistent theory of localization \cite{prl2000, dz, dz2, dz3}.  This implies $\tau(L-z)= B - \tau(z)$ and
\begin{equation}
  2W = \frac{3(\tau_S+\tau_0)}{v_EB^*} \int_0^L dz [B^* - \tau(z)  -2\tau_0 + \tau(z)  ].
\end{equation}
Hence, to leading order,
\begin{equation}\label{friedel}
\langle t \rangle = \frac{3(\tau_S+\tau_0) L}{2v_E}.
\end{equation}
Thus DA yields $\nu =3 (\tau_S+\tau_0)/4$ in Eq.\ (\ref{tnu}). This is scale-independent but depends on exact source depth and extrapolation length $z_0$ which, in turn, is known to depend on internal reflections on the sample surfaces. Without internal reflection at the boundaries $\tau_0=\frac{2}{3}$. We see that DA correctly reproduces the two cases considered within EQRT: the normal incidence has $\tau_S=1$ and the isotropic incidence has $\tau_S=\frac{2}{3}$. For an incidence from direction $\mu_0$ we easily find  $\tau_S=\mu_0$.

\subsection{Delay time for a sphere}

In the following we will obtain the total delay time for a 3D sphere of radius $R$ with equal radiation incident on all points of the outer surface that still may depend on $\mu$. We expect the total delay time---integrated over all outgoing points on the surface---to be independent of angle, so we adopt a spherically symmetric source $F_{in} \delta(r-r_S)/4\pi r^2$. We shall put $w(R+z_0)=0$ as a boundary condition in the diffusion equation
\begin{equation}
    sw(r,s) - \frac{1}{r^2} \frac{d}{dr}\left[ r^2 \frac{v_E}{3}\ell(r)\frac{dw(r,s)}{dr} \right] = F_{in} \frac{\delta(r-r_S)}{4\pi r^2}.
\end{equation}
Clearly, the outgoing flux is normalized, since $-4\pi R^2 \times \left. \frac{1}{3}v_E \ell(R)dw/dr\right|_{r=R, s=0}=F_{in}$. The Friedel identity applies and the delay time is equal to
\begin{eqnarray}
\langle t \rangle = \frac{4\pi}{F_{in}}\int_0^R dr r^2 w(r, s=0).
\end{eqnarray}
We can substitute $dX = -dr / \frac{1}{3}\ell(r)r^2$, with $X(R+z_0)=B <X < X(r=0)=\infty$. For $s=0$ the diffusion equation translates into
\begin{equation}
    -v_E\frac{d^2}{dX^2 }w(X,s=0) = \frac{F_{in}}{4\pi } \delta(X-X_S).
\end{equation}
The solution of this equation satisfying $w(X=B)=0$ and $w(X\rightarrow \infty) < \infty$ to have finite energy in the center of the sphere is
\begin{equation}
   \frac{ 4\pi w(X)}{F_{in}} = -\frac{1}{2v_E}|X-X_S| + \frac{1}{2v_E}(X + X_S) - \frac{B}{v_E}.
   \label{wx}
\end{equation}
The delay time equals
\begin{eqnarray}
\langle t \rangle = \frac{4\pi}{F_{in}}\int_0^{r_S} dr r^2 w(r)+\frac{4\pi}{F_{in}}\int_{r_S}^R dr r^2 w(r).
\label{tsphere1}
\end{eqnarray}
As  follows from Eq.\ (\ref{wx}), for $0<r<r_S$ the energy density $w(r)$ is constant and proportional to  $X_S - B \approx (z_0+z_S) \left. dX/dr \right|_{r=R} = (z_0+z_S)  / [\frac{1}{3} \ell(R) R^2]$. The first term in Eq.\ (\ref{tsphere1}) thus equals $ (z_0+z_S)R/v_E\ell(R)$. The second integral is a surface contribution that is smaller than the first one by a factor $R/z_S$. Thus, for $F_{in} \sim R^2$ we find $W \sim R^3$. This ``normal" scaling implies that
\begin{eqnarray}
\langle t \rangle = \frac{z_S+z_0}{\ell(R)}\frac{R}{v_E}.
\end{eqnarray}
This result is similar to what we have found for the slab. For  normal radiation incident from the far field, and without internal reflection, the front factor equals $1+ 2/3= 5/3$, or equivalently $\nu=1.25$ in Eq. (\ref{tnu}).  For isotropic incident radiation the front factor equals $4/3$, and we recover the universal value $\nu = 1$.

\section{Quasi-1D transport}

In this section we investigate the delay time in a quasi-1D disordered wave guide to see what remains of the universal relation (\ref{tnu}) if the measurement is done only in transmission \emph{or} reflection. This question is extremely relevant for experiments where measuring both transmission and reflection may be problematic. We know that, in principle, in a quasi-1D wave guide with $N$ transverse modes all waves are localized with a localization length $\xi \sim N \ell_B$ \cite{carlo}. Here $\ell_B$ is the transport mean free path in the absence of Anderson localization effects. In the self-consistent theory of localization, the quasi-1D geometry is described by the diffusion equation (\ref{da2}) with $q=0$, supplemented by a self-consistent equation for the position-dependent transport mean free path \cite{dz,note1}:
\begin{eqnarray}\label{eqstau}
\frac{1}{\ell(\tau, s)}= \frac{d\tau}{dz}  = \frac{1}{\ell_B} + \frac{1}{3\xi}\, G(\tau ,\tau,s),
\end{eqnarray}
which also depends on $s$.
The Green's function of Eq.\ (\ref{da2}) is given by Eq.\ (\ref{Gserie}). For $\tau \ll B$, i.e. for a semi-infinite wave guide, the sum in Eq.\ (\ref{Gserie}) is essentially an integral and $G(\tau, \tau) = 3 (\tau+\tau_0)$. We then easily find $\ell(z,s=0)=[\ell_B/(1+z_0/\xi)]  \exp(-z/\xi)$.

The dynamics can be included using standard perturbation theory and treating $s \ell(\tau, s) / v_E$ as a small perturbation in Eq.\ (\ref{da2}). As in the previous section, we account for boundary conditions by introducing an extrapolation length $z_0$, $B^*= B+2\tau_0$ and $\tau^*=\tau+ \tau_0$. Since $\ell(\tau,s)$ is suppressed by localization and since $s$ is supposed to be a small hydrodynamic frequency, the perturbation can be argued  to be``small". In  first-order perturbation theory the eigenvalues of the diffusion equation change by $s W_{nn}/v_E$ with
\begin{equation}\label{delta}
W_{nm}  \equiv \langle \Phi_n | {\ell(\tau, s)} |\Phi_m \rangle
\end{equation}
and the eigenfunctions by
\begin{equation}\label{eigenf}
  \delta \Phi_n(\tau)=-\frac{ 3\sqrt{2} s}{\sqrt{B^*}v_E} \sum_{m \ne n}^\infty \frac{W_{nm}}{ q_m^2 - q_n^2}
   \sin(q_m\tau^*).
\end{equation}
Hence,
\begin{equation}\label{GOmega}
  G(\tau, \tau',s)=  \sum_{n=1}^\infty \frac{\Phi_n(\tau^*,s)\Phi_n(\tau'^*,s) }{s W_{nn}(s) /v_E +  \frac{1}{3}  q^2_n}.
\end{equation}

We easily find that
\begin{eqnarray}\label{delta2}
W_{nm} &=& \ell_B \frac{2}{B^*}\int_{-\tau_0}^{B+\tau_0} d\tau \frac{ \sin (\pi n \tau^*/B^*)\sin (\pi m \tau^* /B^*)}{1+ (\tau^*\ell_B/\xi) (1-\tau^*/B^*)}  \nonumber \\
\, &=& \frac{2\xi}{B^*}\int_0^1 dx \frac{ \sin (\pi n x)\sin (\pi m x)}{\xi/\ell_B B^*+ x(1-x)}.
\end{eqnarray}
$W_{nm}$ is intrinsically a parameter of the dynamics, and it is of no relevance when $s=0$.
 We can map the denominator in the sum of Eq.\ (\ref{GOmega}) onto the familiar diffuse  $1/[ s \ell_B /v_E + \frac{1}{3} (\pi n\ell_B /L)^2]$. It is tempting to conclude that  each diffusion mode has now achieved its own diffusion constant
 $D_n(s=0) = \frac{1}{3} (v_E \ell_B/W_{nn}) L^2/\ell_B^2 B^2$. We could proceed by saying that the transport velocity is affected by the dynamic kernel $W_{nn}/\ell_B \sim \xi/ B <1 $ according to $v_E^{(n)}\sim v_E / W_{nn}$ and would thus be \emph{enhanced} by the scale-dependent diffusion. In this logic, the  transport velocity would be  associated with the energy density $\widetilde{W }\sim \ell(\tau) W$ which is the conserved quantity featuring in the diffusion equation (\ref{da2}). Such a definition of $v_E$ is based on long-time tails of energy density in contrast to Eq.\ (\ref{vE}) that relies on the average delay time and the \emph{genuine} energy density $W$ which scales normally as $W \sim V$ even in the presence of scale-dependent diffusion. Apparently, the transport velocity is no longer uniquely defined when localization effects come into play. A deeper analysis is clearly necessary here but it is beyond the scope of this work.

\subsection{Delay time in reflection}

We are interested in calculating the weighted delay time in reflection: $\langle R(t) t \rangle = \langle R(\omega) d\phi/d\omega \rangle = -dR(s)/ds$ at $s=0$, with $R(s)$ being the Laplace transform of the average reflection coefficient $R(t)$. In the diffuse regime, it is easy to see that $\langle R(t) t \rangle \propto L/v_E$ up to a factor related to boundary effects, which provides  a unique opportunity to measure the transport velocity directly. For normal diffusion, the energy density decays essentially algebraically between the two boundaries of the medium, and the total energy is thus proportional to $L$. In the following we ignore the $s$-dependence of $\ell$ imposed by localization effects, but we shall perform numerical calculations to make a comparison.

\begin{figure}
\includegraphics[width=0.99\columnwidth]{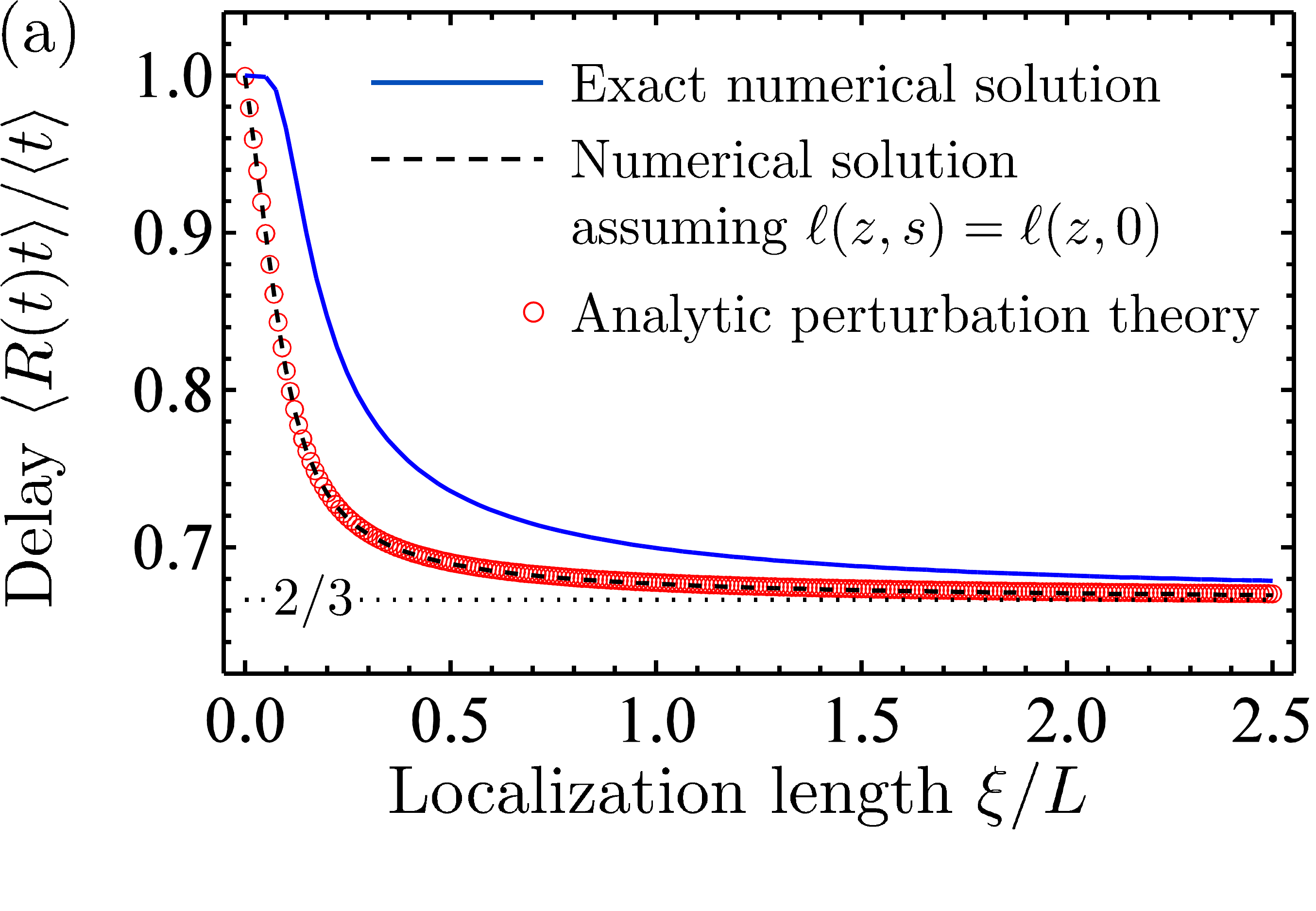}\\
\vspace{-4mm}
\includegraphics[width=0.99\columnwidth]{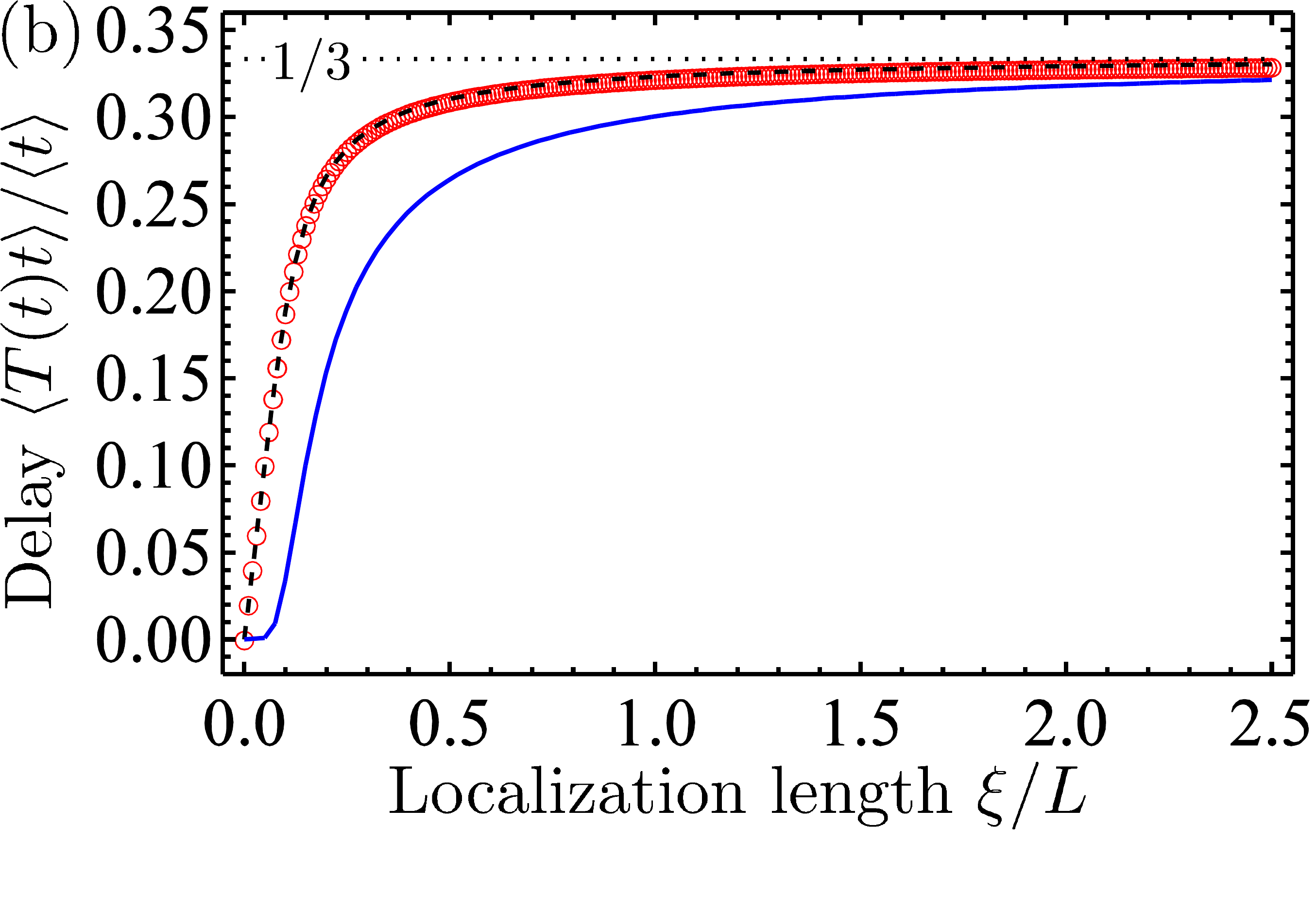}\\
\vspace{-6mm}
\caption{Weighted time delay in reflection (a) and transmission (b) of a wave through a quasi-1D disordered wave guide as a function of the ratio of localization length $\xi$ to the length $L$ of the wave guide. Blue solid lines show the results following from the exact numerical solution of the self-consistent equations (\ref{da}) and (\ref{eqstau}), black dashed lines show the numerical solutions in which the dependence of $\ell(z,s)$ on $s$ in neglected. The latter coincide with our perturbation theory results (\ref{Tr1D}) and (\ref{delayT}) shown by red circles. We used $L = 200 \ell_B$, $z_0 = 0$, and $z_S = \ell_B$ for this figure.}
\label{fig_delay}
\end{figure}

We consider a perfect point source $ \delta(\tau - \tau_S)$ in Eq.\ (\ref{da2}). The average total reflection coefficient (integrated over all angles $-1 < \mu < 0$) is
\begin{equation}\label{phaseref}
R(s) = +\frac{1}{3} \partial_\tau G(\tau=0, \tau_S, s).
\end{equation}
 It can easily be seen that
 \begin{eqnarray}
 R(0) &=& \frac{2}{B^*} \sum_{n=1}^\infty \frac{\cos(q_n \tau_0)\sin (q_n \tau_S^*)}{q_n} \\
   \, &\approx &  \frac{2 }{\pi } \sum_{n=1}^\infty \frac{\sin (n \pi x)}{n} = 1-x
 \end{eqnarray}
 with $x= (\tau_S+\tau_0)/B^*$ the  transmission, exponentially small in the localized regime, and for a source near $z = 0$. We have two contributions to the delay time $\langle R(t) t \rangle =-\partial_s R(s=0)$. The first comes from the modified eigenvalues:
\begin{eqnarray}\label{delayR2}
\langle R(t) t \rangle^{(1)} &=& \frac{2}{B^* \frac{1}{3}v_E } \sum_{n=1}^\infty W_{nn} \frac{\sin(q_n \tau_S^*)}{q^3_n} \nonumber \\
     \, &=& \frac{6(B^*)^2}{v_E} \frac{1}{\pi^3} \sum_{n=1}^\infty W_{nn} \frac{\sin (\pi n x)}{n^3 }\nonumber \\ &\approx& \frac{6(B^*)^2}{v_E}\frac{ x }{\pi^2} \sum_{n=1}^\infty  \frac{W_{nn}}{n^2}.
\end{eqnarray}
The second contribution to the delay time stems from the modified eigenfunctions, which generate
\begin{eqnarray}
  \langle R(t) t \rangle^{(2)} &=& \frac{6}{B^* v_E }\sum_{n=1}^\infty \sum_{m \ne n}^\infty \frac{q_m W_{nm}}{q_m^2-q_n^2} \frac{\sin (q_n \tau_S^*)}{q_n^2} \nonumber \\
  \,  &+ &  \frac{6}{B^* v_E} \sum_{n=1}^\infty \sum_{m \ne n}^\infty \frac{W_{nm}}{q_m^2-q_n^2} \frac{\sin (q_m \tau_S^*)}{q_n}.
\label{eq1}
\end{eqnarray}
Upon interchanging $n$ and $m$ in the first line we rewrite Eq.\ (\ref{eq1}) as
\begin{eqnarray}
  \langle R(t) t \rangle^{(2)} &=& \frac{6}{B^* v_E }\sum_{n=1}^\infty \sum_{m \ne n}^\infty \frac{W_{nm}}{q_m^2- q_n^2} \sin (q_m \tau_S^*)
  \nonumber \\
  &\times& \left( {\frac{1}{q_n}- \frac{q_n}{q_m^2}} \right) \nonumber\\
     \, &=&  \frac{6(B^*)^2}{v_E} \frac{1}{\pi^3}
  \sum_{n=1}^\infty \sum_{m \ne n}^\infty \frac{W_{nm}}{m^2n} {\sin (\pi m \tau_S^*/B^* )} \nonumber \\
  &\approx& \frac{6(B^*)^2}{v_E} \frac{\tau_S^*}{B^* \pi^2}\sum_{n=1}^\infty \sum_{m \ne n}^\infty \frac{W_{nm}}{mn} \nonumber \\
  &=& \frac{6B^*\tau_S^*}{v_E} \frac{1}{\pi^2}\sum_{n=1}^\infty \left(\sum_{m \ne n}^\infty \frac{W_{nm}}{mn} - \frac{W_{nn}}{n^2} \right) \nonumber \\
  &=& - \langle R(t) t \rangle^{(1)} \nonumber \\
  &+& \frac{3\tau_S^*\xi}{v_E} \int_0^1 dx \frac{(1-x)^2 }{\xi/\ell_B B^* +x(1-x)}.
\end{eqnarray}
The approximation assumes $\tau_S,\tau_0  \ll B$.  We can interchange $x$ and $1-x$ and write $ (1-x)^2 + x^2 = -2x(1-x) +1$ to get
 \begin{eqnarray}
  \langle R(t) t \rangle &=& -\langle T(t) t \rangle
  \nonumber \\
  &+& \frac{3\tau_S^*\xi}{2v_E} \int_0^1 dx \left( \frac{B^*\ell_B/\xi}{1 +(B^*\ell_B/\xi) x(1-x)} \right)  \nonumber \\
  &=&  -\langle T(t) t \rangle + \frac{3\tau_S^*\ell_B}{2v_E} \int_{-z_0}^{L+z_0} dz,
\end{eqnarray}
where the delay in transmission $\langle T(t) t \rangle=3\xi\tau_S^*(1-L/B\ell_B)/v_E$  is obtained in the next section. Anticipating this result gives us
\begin{equation}\label{Tr1D}
   \langle R(t) t \rangle = \frac{3(\tau_S+\tau_0)L}{2v_E} \left(1 -\frac{2\xi}{L} + 2\frac{\xi}{B \ell_B} \right),
\end{equation}
where all corrections of order $z_0/L $ have been ignored.
This expression approaches $(\tau_S+\tau_0)L/v_E $  in the diffuse regime $B \ll\xi $, and converges to $ \frac{3}{2} (\tau_S+\tau_0)L/v_E $ deep in the localized regime. Contrary to the total delay, the delay time in reflection varies, though  little, upon going from the diffuse into the localized regime. We illustrate this in Fig.\ \ref{fig_delay}(a) where results of different approaches to the calculation of $\langle R(t) t \rangle$ are compared. We see, in particular, that our perturbational result (\ref{Tr1D}) is not exact and corresponds to a solution assuming $\ell(z,s) = \ell(z, 0)$. Its dependence on the strength of localization effects quantified by the ratio $\xi /L$ is, however, similar to the one exhibited by the exact solution of Eqs.\ (\ref{da}) and (\ref{eqstau}).

\begin{figure}
\includegraphics[width=0.99\columnwidth]{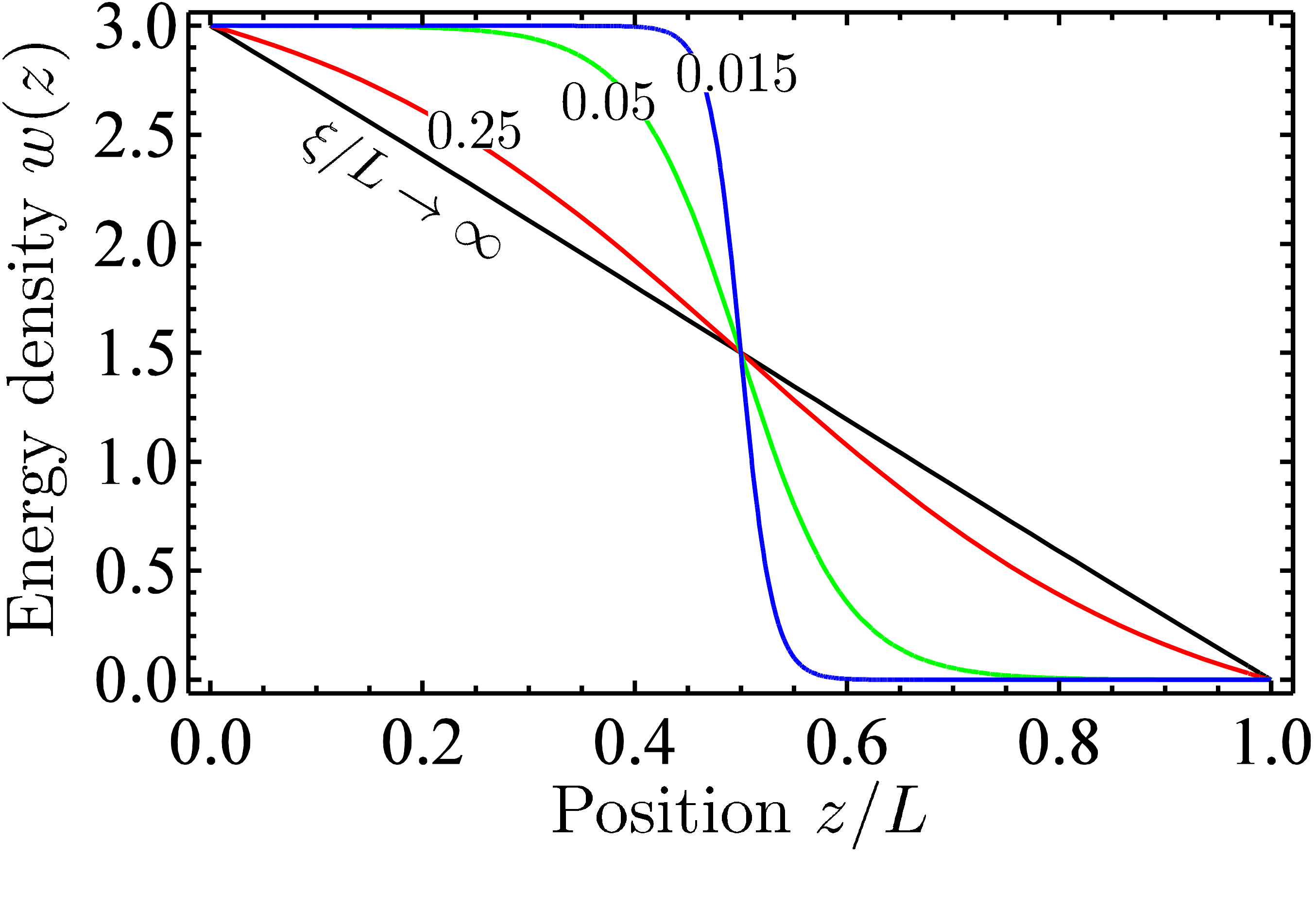}\\
\vspace{-4mm}
\caption{Energy density, in units of $F^+/v_E$ (with $F^+$ the incident flux) inside a disordered wave guide for a wave incident at $z=0$, as a function of position $z$ for four different values of the ratio of localization length $\xi$ to the length $L$ of the wave guide. We used $z_0 = 0$ and $z_S = \ell_B$ for this figure.
Independent of $\xi/L$, the total energy $W$ inside the wave guide is always equal to $3(\tau_S + \tau_0)L F^+/2 v_E  = 3L F^+/2 v_E$.
}

\label{fig_w}
\end{figure}

A naive argument would suggest that in the localized regime, a wave penetrates only a distance $\xi$ into the medium. This would lead to a much shorter delay time of order $\xi/v_E$ in reflection. This argument is apparently wrong by a factor $L/\xi$, because the energy does not decay as $\exp(-z/\xi)$. The lack of a penetration depth $\xi$ in the energy density $w(z)$ is seen when we translate the $(1-\tau/B)-$profile of the energy density back to the $z$ variable.  We observe that $w(z)$, in the localized regime, is actually \emph{flat} on both sides of the medium with a steep descent in a region of size $\xi$ around $z=L/2$. This is illustrated in Fig.\ \ref{fig_w} where we show profiles $w(z)$ for several values of $\xi/L$. We observe that $w(z)$ evolves from a linear decay in the diffuse regime $\xi/L \to \infty$ to a step-like function deep in the localized regime $\xi/L \ll 1$. For a given incident flux density $F^+$, the energy density is always the same in the middle of the slab, whatever $\xi/L$. As counterintuitive as this can appear, the naive assumption of initial decay of $w(z)$ over a region of size $\xi$ is not confirmed by the $z$-dependent description of localization, since near the boundaries the waves are not localized. However, for a source in the middle of the wave guide, we find $w(\tau) \sim |\tau-\frac{1}{2}B| - \frac{1}{2}B $. The energy density as a function of $z$ is then
\begin{equation}
  w(z)\sim e^{L/2\xi}e^{-|z-L/2|/2\xi} -1,
\end{equation}
i.e. $w(z)$ decays exponentially on both sides from the source and the waves are localized deep in the sample.
The delay time, on both sides, then varies as $\langle t \rangle_{R,T} = L^2/D_B$  for $L \ll \xi$, and grows exponentially as $\exp (L/2\xi) $ deep in the localized regime.

\subsection{Delay time in transmission}

The transmission is given by
\begin{equation}\label{phasetrans}
  T(s) = -\frac{1}{3} \partial_\tau G(\tau=B, \tau_S, s)
\end{equation}
and we easily obtain
\begin{eqnarray}
   T(0) &=& -\frac{2}{B^*} \sum_{n=1}^\infty \cos(n\pi B/B^*)\frac{\sin (q_n \tau_S^*)}{q_n} \nonumber \\
   \, &\approx&  \frac{2 }{\pi } \sum_{n=1}^\infty (-1)^{n+1}\frac{\sin (n \pi x)}{n} = x,
 \end{eqnarray}
where we recall that $x=\tau_S^*/B^*$. Similar to Eq. (13) we  find the first contribution to delay time in transmission,
\begin{eqnarray}\label{delayT2}
\langle T(t) t \rangle^{(1)} &=& \frac{2}{B^* \frac{1}{3}v_E } \sum_{n=1}^\infty (-1)^{n+1} W_{nn} \frac{\sin(q_n \tau_S^*)}{q^3_n} \nonumber \\
&\approx& \frac{6(B^*)^2}{v_E}\frac{ x }{\pi^2} \sum_{n=1}^\infty  (-1)^{n+1} \frac{W_{nn}}{n^2}.
\end{eqnarray}
The second contribution is
\begin{eqnarray}
  \langle T(t) t \rangle^{(2)} &=& -\frac{6}{B^* v_E }\sum_{n=1}^\infty \sum_{m' =1}^\infty \frac{q_m W_{nm}}{q_m^2-q_n^2} \frac{\cos(m\pi)\sin (q_n \tau_S^*)}{q_n^2} \nonumber\\
  \,  &- &  \frac{6}{B^* v_E} \sum_{n=1}^\infty \sum_{m \ne n}^\infty \frac{W_{nm}}{q_m^2-q_n^2} \frac{\cos(n\pi) \sin (q_m \tau_S^*)}{q_n}. \nonumber \\
\end{eqnarray}
Again upon interchanging $n$ and $m$ in the first term, we obtain
\begin{eqnarray}
  \langle T(t) t \rangle^{(2)} &=& \frac{6}{B^* v_E }\sum_{n=1}^\infty \sum_{m \ne n}^\infty \frac{(-1)^{n+1}W_{nm}}{q_m^2- q_n^2} \sin(q_m\tau_S^*)
\nonumber \\
&\times& \left( {\frac{1}{q_n}- \frac{q_n}{q_m^2}} \right) \nonumber \\
   \, &= &   \frac{6B^*\tau_S^*}{v_E} \frac{1}{\pi^2}\sum_{n=1}^\infty (-1)^{n+1} \nonumber \\
   &\times&
   \left(\sum_{m=1}^\infty \frac{W_{nm}}{m n} - \frac{W_{nn}}{n^2} \right) \nonumber \\
  &=& -\langle T(t) t \rangle^{(1)} \nonumber \\
  &+& \frac{6B^*\tau_S^*}{v_E} \frac{1}{\pi^2}\sum_{n=1}^\infty\sum_{m=1}^\infty  (-1)^{n+1}\frac{W_{nm}}{m n}.
\end{eqnarray}
Hence, the mean weighted delay time is transmission is
\begin{eqnarray}\label{delayT}
  \langle T(t) t \rangle  &= & \frac{6B^*\tau_S^*}{v_E} \frac{1}{\pi^2}\sum_{n=1}^\infty \sum_{m=1}^\infty(-1)^{n+1}  \frac{W_{nm}}{m n} \nonumber \\
  &=& \frac{3\xi \tau_S^* }{v_E B^*} \int_0^{B^*} d\tau \left[ 1-\frac{1}{1 + (\ell_B/\xi) \tau(1-\tau/B^*)}\right]\nonumber \\
  &=& \frac{3\xi \tau_S^* }{v_E } \left( 1 - \frac{1}{B^*\ell_B}\int_{0}^{L+2z_0} dz\right)
  \nonumber \\
  &\approx& \frac{3\xi \tau_S^* }{v_E}\left( 1-\frac{L}{B\ell_B}\right).
   \end{eqnarray}
For $\xi \gg B\ell_B$ (diffuse regime), $\langle T(t) t \rangle = (\tau_S+\tau_0)L/2v_E $ and thus the un-weighted delay time is $\langle t \rangle_T = L^2/6D_B$. Upon entering the localized regime ($\xi < B\ell_B$), this value saturates  exponentially towards the $L$-independent value $\langle T(t) t \rangle =  3(\tau_S+\tau_0)\xi/v_E$. The un-weighted delay time is now equal to $\langle t \rangle_T = 3\xi B/v_E$. Naively, we could have expected $\langle t \rangle_T = L^2/6D$ with a reduced ``scale-dependent" diffusion constant $D = D_B L/B\ell_B$.  This argument turns out to be wrong by a factor of order $L/\xi$. The dependence of $\langle T(t) t \rangle$ on $\xi/L$ is illustrated in Fig.\ \ref{fig_delay}(b). Similarly to the case of reflection, we observe deviations of Eq.\ (\ref{delayT}) from the exact numerical calculation, that also takes into account the dependence of $\ell$ on the dynamical parameter $s$. Quite remarkably, even though our results for both $\langle R(t) t \rangle$ and $\langle T(t) t \rangle$ are only approximate, their sum $\langle t \rangle = \langle R(t) t \rangle + \langle T(t) t \rangle$ is equal to the total delay time (\ref{normal2}) \emph{exactly}.

\section{3D slab}

In this section we use the perturbation theory to study the delay time in a 3D slab. We also calculate other properties characteristic for 3D media: coherent backscattering, transverse spreading of a focused incident beam, and time-reversal focusing in the localized regime. As in section \ref{friedelsec}, we consider a slab of disordered medium confined between planes $z=0$ and $z=L$.

\subsection{Delay time}

The results obtained above for a quasi-1D system can be used---\emph{mutatis mutandis}---for the 3D slab geometry, provided  that we integrate over the transverse surface (hence $\mathbf{q}=0$). Physically this corresponds to measuring the delay times upon integration over the entire boundary surfaces.  It can easily be seen that the total weighted delay in reflection, for example, is $\langle R(t) t \rangle = - d R(s=0, \mathbf{q}=0)/ds$. However, the diffusion coefficient is no longer determined by the return probability $\tau(1-\tau/B)$ formula valid in a quasi-1D wave guide. For $\mathbf{q}=0$, Eq.\ (\ref{da}) remains valid and $\tau $ can be defined similarly in terms of $\ell(z)$. We define
\begin{eqnarray}\label{delta3}
W_{nm} &=& \frac{2}{B}\int_0^B d\tau \ell(\tau)\sin (\pi n \tau/B)\sin (\pi m \tau /B)  \nonumber \\
\, &=& 2\int_0^1 dx  \sin (\pi n x)\sin (\pi m x) \ell(\tau=xB)
\end{eqnarray}
and find
\begin{equation*}
\langle t \rangle_R \approx  \langle R(t) t \rangle = \frac{3B\tau_S^*}{v_E} \int_0^1 dx (1-x)^2 \ell(\tau=xB).
\end{equation*}

Let us consider a profile $\ell(z)= \ell_B/(1 + z/\xi_c)$ for $0 < z < L/2$ (and mirrored on the other side of $z = L/2$), typically valid at the 3D mobility edge \cite{prl2000}. Since $ \ell_B d\tau = dz (1+ z/\xi_c)$, we find
\begin{equation}
 \langle R(t) t \rangle = \frac{3B\ell_B}{v_E} \int_0^{1/2} dx \frac{(1-x)^2+x^2 }{\sqrt{1 + (B/\tau_{\xi_c})x }}
\end{equation}
with $\tau_{\xi_c} = \xi_c/\ell_B$ and $B= L/\ell_B+ (L/2)^2/\ell_B\xi_c$. For $L \ll \xi_c $ we obtain $\langle R(t) t \rangle = (\tau_S+\tau_0)L/v_E$. For $L \gg \xi_c$, we have $\langle R(t) t \rangle = 3(\tau_S+\tau_0) \frac{23}{60} (L^2/4\xi_c v_E) \times 2\xi_c /L $. The correlation length $\xi_c$ drops out and $\langle R(t) t \rangle \rightarrow \frac{23}{20} \, L/v_E$ scales with the sample size $L$. The front factor $1.15 $ is smaller than the factor $1.5$ obtained for the localized regime, and slightly exceeds the value $1$ in the diffuse regime.

In transmission we proceed in a similar way and obtain
\begin{equation}
 \langle T(t) t \rangle = \frac{3B\tau_S^*}{v_E } \int_0^{1/2} dx 2x(1-x) \ell(\tau=xB).
\end{equation}
This yields in the localized regime $\langle T(t) t \rangle \rightarrow \frac{7}{20} \, (\tau_0+\tau_s)L/v_E $, smaller than but of the same order of magnitude as the weighted delay in reflection. Note that the Friedel identity (\ref{friedel}) is  obeyed ($7/20 + 23/20 = 3/2)$. Again, the $s-$dependence affects both delays, but not their sum. In the diffuse regime, the ratio of weighted delays in reflection and transmission is $2:1$, at the mobility edge this is close to $3:1$. In the localized regime the ratio further grows up to $L/\xi$.

\subsection{Transverse diffusion}

In Ref.\ \cite{john}, dynamic transverse diffusion was used as a probe for Anderson localization. In the following we will  treat $\frac{1}{3} \ell(\tau,s)^2 q^2$ as a small perturbation in Eq.\ (\ref{da2}) and study stationary properties. The Green function of the diffusion equation (\ref{da2}) is most conveniently written as
\begin{equation}\label{diffq}
    G(\tau,\tau', \mathbf{q}, s) =\sum_n \frac{\Phi_n(\tau,q,s) \Phi_n(\tau',q,s)}{\frac{s}{v_E}W_{nn} + \frac{1}{3} V_{nn} q^2 + \frac{1}{3}  q_n^2},
\end{equation}
where we have introduced
\begin{equation}
    V_{nm} = \langle \Phi_n | \ell(\tau,s)^2 |\Phi_m \rangle.
\end{equation}
Note that the Green function (\ref{diffq}) looks as if \emph{anisotropic diffusion processes different for each mode} were at work. The modification of eigenfunctions due to the perturbation $\frac{1}{3} \ell(\tau,s)^2 q^2$ is
\begin{equation}\label{eigenf2}
  \delta \Phi_n(\tau,\mathbf{q})=-\frac{ \sqrt{2} q^2}{\sqrt{B^*}} \sum_{m \ne n}^\infty \frac{V_{nm}}{ q_m^2 - q_n^2}
   \sin(q_m\tau^* ).
\end{equation}

Given a stationary source of waves of small transverse size at $\{ x_S = y_S = 0, z_S \sim \ell_B \}$, the stationary energy at depth $z$ and transverse distance $\mathbf{R}$ is $\rho(z,\mathbf{R})= G(z,z'= \ell_B,\mathbf{R},s=0)$, with transverse Fourier transform $\rho(z,q)$. The mean transverse energy spread at a depth $z$ can be quantified by $\langle R(z)^2 \rangle = \langle \rho(z, \mathbf{R})\mathbf{R}^2\rangle /\langle \rho(z, \mathbf{R})\rangle = -(1/q) \partial_q[ q \partial_q \rho(z,q)]/ \rho(z,q)$ at $q=0$ (we use $\mathbf{R} = \{x, y \}$). Hence we need to expand to order $q^2$. This expansion is similar to the expansion in $s$ used to find the delay time, and we can copy the result (\ref{delayT}) \emph{mutatis mutandis},
\begin{eqnarray}
 \delta \rho(\tau,q) &=& -6 \frac{q^2}{B}\sum_{n=1}^\infty \sum_{m=1}^\infty \frac{V_{nm}}{q_n^2q_m^2} \sin q_n\tau^* \sin q_m\tau_S^* \nonumber \\
 &=&  \frac{-12B^2q^2}{\pi^4}\int_{-\tau_0}^{B+\tau_0} d\tau' \nonumber \\
 &\times& \sum_{n=1}^\infty \sum_{m=1}^\infty \frac{\sin q_n \tau^* \sin q_m \tau_S^* \sin q_n \tau'^* \sin q_m \tau'^*}{m^2n^2}
 \nonumber \\
 &\approx& -3q^2B\tau_S^*
 \int_0^{B^*} d\tau'  \ell(\tau'-\tau_0)^2 \nonumber \\
 &\times& \left[\frac{\min(\tau,\tau')}{B^*} -\frac{\tau\tau'}{{B^*}^2}\right] \left(1-\frac{\tau'}{B^*}\right).
\end{eqnarray}
Since $\rho(\tau,q=0) = 3\tau_S^*(1-\tau^*/B^*)$ we find
\begin{eqnarray}\label{transverses0}
\langle R^2(\tau)\rangle= 4  \int_0^{\tau^*} d\tau' \ell(\tau'-z_0)^2 \tau'\left(1-\frac{\tau'}{B^*}\right) \nonumber \\
+ \frac{\tau^*}{B^*-\tau^*} \int_{\tau^*}^{B^*} d\tau' \ell(\tau'-z_0)^2\left(1-\frac{\tau'}{B^*}\right)^2.
\end{eqnarray}

Near $\tau=B$ (in transmission) the second term is negligible, and we obtain
 \begin{eqnarray}\label{transversesT}
\langle R^2(L)\rangle= 4 \int_0^{B^*} d\tau'  \frac{\ell(\tau'-z_0)^2}{\ell_B^2}\tau'\left(1-\frac{\tau'}{B^*}\right).
\end{eqnarray}
In the diffuse regime $\ell= \ell_B$ and $\tau = z/\ell_B$ so that $\langle R^2(L) \rangle =2L^2/3$. In the localized regime, $\ell(z)$ can be approximated by its profile in quasi-1D: $\ell_B/\ell = 1+ (\tau/\tau_\xi) (1-\tau/B)$, so that
\begin{equation}\label{transverses0loc}
\langle R^2(L)\rangle= 4 \xi \int_0^L dz  \left( 1- \frac{\ell(z)}{\ell_B}\right) \approx 4\xi L.
\end{equation}
These findings agree with previous results \cite{nicolas, note1}.  At a given optical depth $\tau\ll B$ we can take the limit of the half-space ($L \to \infty$) to see that
\begin{equation}\label{transverses0half}
\langle R^2(z)\rangle= 4 \xi^2 \log\left( 1+ \tau^*/\tau_\xi\right) = 4 \xi (z+z_0)
\end{equation}
and, in particular, $\langle R^2(0)\rangle= 4 \xi z_0 $ in reflection. This result is surprisingly simple: the mean-square size of the transverse region in which the wave energy is concentrated grows \emph{linearly} with the depth into the medium. For normal diffusion  we also find a linear growth  $\langle R^2(z)\rangle= \frac{4}{3}L(z+z_0)(1- z/2L)$ with, however, a much larger slope that even diverges for a half-space since the transverse profile then becomes algebraic.

\subsection{Coherent backscattering and time reversal}

In the diffusion approximation and for normal incidence, the stationary CBS profile is approximately given by \cite{akk}
\begin{equation}\label{CBS}
    C(\mathbf{Q}) = G(\tau=1, \tau_S , s=0 , \mathbf{q}=\mathbf{Q})
\end{equation}
with $\mathbf{Q}=\mathbf{k}+\mathbf{k}'$ that vanishes at exact backscattering and $G$ given by Eq.\ (\ref{diffq}). For $\tau_S \approx 1$, this yields the familiar formula for CBS of a normally incident plane wave. For $\tau_S$ somewhere inside the slab,  this expression actually describes the ensemble-averaged time-reversed profile by a perfect pointlike time-reversal machine at an optical depth $\tau_S$ \cite{mathias, bartprl}.   Using the perturbational approach of the previous section, we find  the emerging specific intensity to be
\begin{eqnarray}
  \delta C(\mathbf{Q}) &=& -\frac{6Q^2B^2}{\pi^3}(1+\tau_0) \sum_{n,m} \frac{V_{nm}}{nm}\sin(q_n\tau_s^*).
\end{eqnarray}
Recalling the definition of $V_{nm}$, we simplify this to
\begin{eqnarray}
  \delta C(\mathbf{Q}) &=&
  -3Q^2 (1+\tau_0)\int_0^{B^*} d\tau \ell(\tau-\tau_0)^2 \nonumber \\
  &\times& \left[ \min(\tau,\tau_S)-\frac{\tau_S\tau}{B^*} \right] \left(1- \frac{\tau}{B^*}\right).
  \end{eqnarray}
The background is given by
\begin{eqnarray}
    R(0) &=&\frac{6}{B}\sum_n \frac{\sin q_n(1+\tau_0) \sin q_n\tau_S^*}{q_n^2} \\
    &=& {3} (1 + \tau_0)(1-\tau_S/B),
\end{eqnarray}
so that the normalized CBS profile becomes ($\tau_S > 1$)
\begin{eqnarray}\label{cone}
    C(Q) &=& 1 -Q^2 \frac{1}{1-\tau_S/B }\int_0^{B^*} d\tau \ell(\tau-z_0)^2
    \nonumber\\
  &\times&  \left[ \min(\tau, \tau_S) -\frac{\tau\tau_S}{B^*}\right] \left( 1 -\frac{\tau}{B^*}\right).
\end{eqnarray}

This result can be discussed in various limits. In the weak-disorder limit $\ell(z) = \ell_B$ and $B=L/\ell_B$. Hence, the rounding of the CBS cone is typically $-Q^2 \tau_S \ell_B L$, i.e. it is rounded due to finite-size effects. For a half-space ($L \to \infty$) this result is of little interest since the line profile is known to turn into the familiar cusp $-|Q|\tau_S \ell_B$, which is beyond the present perturbation theory as  $Q^2$ has been assumed to be the leading small parameter. However, in the localized regime the limit of $L \to \infty$ can be taken since the integral in Eq.\ (\ref{cone}) converges and we obtain
 \begin{equation}\label{coneloc}
    C(Q) = 1 - 4Q^2\xi (z_S+z_0).
 \end{equation}
For CBS of a plane wave at normal incidence $z_S=\ell_B$ and we recover the rounding proportional to $-Q^2 \ell_B \xi$ as predicted previously \cite{prl2000}.

Finally, for a time-reversal experiment with a time-reversal machine at depth $z_S$ we see, quite surprisingly, that the angular size of the focal spot $\delta \theta$ narrows down with (the \emph{genuine}) depth according to $\delta \theta \propto 1/ k \sqrt{\xi z_s}$ in contrast to $\delta \theta \propto 1/ k z_s$ in the diffuse regime. We could have expected $\delta \theta \sim 1/ k \xi$, and arguably an impossibility to time-reverse well as $z_S > \xi$, but this turns out to be a wrong expectation. Note, however, that we do not expect \emph{auto}-focusing (i.e. focusing in the absence of ensemble averaging) to occur in the localized regime, because strong and long-range correlations should prevent self-averaging of a signal with even a relatively large bandwidth. A full discussion of this issue is beyond the scope of the present work.

\section{Conclusions}

We have presented very general arguments based on the phenomenological equation of radiative transfer to derive simple expressions for the average delay time of a wave in a disordered medium. Our reasonings apply in all regimes of wave scattering, including the regime of strong (Anderson) localization. Specific examples of delay time calculations are provided for different geometries (a slab or a sphere) and different incident waves (an isotropic source or an incident plane wave). Detailed considerations of wave dynamics allowed us to suggest definitions for the energy transport velocity in the localized regime and to demonstrate that a unique definition for the latter may be difficult to achieve. In addition, we develop a novel perturbational approach to radiative transfer of localized waves in quasi-1D and 3D disordered media. This has enabled us to calculate the delay time measured separately in transmission \emph{or} reflection which may be important to design experiments. We also apply our perturbation theory to study how well-known mesoscopic phenomena such as the transverse spreading of a wave packet,  coherent backscattering, and  time-reversal are affected by Anderson localization effects. A future study may be devoted to calculation of time-dependent quantities, such as the time-dependent transmission and reflection coefficients, in the framework of our perturbational approach.

\section*{Acknowledgements}
We would like to thank Roger Maynard for his continuous interest and support of this and related works. We thank R\'{e}mi Carminati and Romain Pierrat for helpful discussions about the delay time.
This work is supported by the Agence Nationale de la Recherche under grant ANR-14-CE26-0032 LOVE, and by the PICS program of the CNRS (project Ultra-ALT).


\begin{thebibliography}{16}
\bibitem{note1} Our definition of the localization length $\xi$ differs from that of Refs.\ \cite{dz,dz2,dz3} by a factor of 2.
\end{thebibliography}
\end{document}